\documentclass[a4paper,11pt]{article}
\usepackage{jinstpub} 
\usepackage{lineno}
\usepackage{subcaption}

\title{\boldmath TCP/IP based Remote Firmware Upgradation for INO ICAL RPC-DAQ Modules}

\author[a,1]{Yuvaraj Elangovan,\note{Corresponding author.}}
\author[a]{Mandar Saraf,}
\author[a]{B. Satyanarayana,} 
\author[a]{S.S. Upadhya,} 
\author[a]{Nagaraj Panyam,} 
\author[a]{Ravindra Shinde,} 
\author[a]{Gobinda Majumder,} 
\author[a]{D. Sil,} 
\author[a]{Pathaleswar,} 
\author[a]{Aditya Deodhar} 
\author[a]{and K.C. Ravindran,} 
\affiliation[a]{Tata Institute of Fundamental Research, Mumbai, India}

\emailAdd{yue8@pitt.edu}

\abstract{
The INO ICAL (India-based Neutrino Observatory Iron Calorimeter) experiment is an upcoming mega-science project currently in the developmental stages. This initiative employs over 28,800 Resistive Plate Chambers (RPC) based charged particle detectors used for tracking muon events. Each of these detectors incorporates an FPGA-based Digital Front End known as RPC-DAQ, with the primary objective of measuring the position and timing of particle interactions within the respective RPCs. The firmware embedded in the RPC-DAQ FPGA is designed to support this logic. The ICAL experiment is a 50-kiloton iron structure arranged in a stacked geometry where the RPCs along with their associated electronics are positioned between these iron plates. Reprogramming individual or groups of RPC-DAQs proves to be a challenging and time consuming task. To address the complexity of upgrading firmware for these typically inaccessible RPC-DAQs this paper introduces an innovative approach that utilizes the existing Ethernet interface, employed for command transmission and data acquisition to upload firmware. A customized handshaking architecture has been designed using the TCP protocol for this experiment. The firmware binary file is segmented into TCP packets and transmitted over Ethernet. The soft-core processor instantiated in the RPC-DAQ FPGA receives these firmware packets overwriting the existing firmware in the flash memory. Upon rebooting the RPC-DAQ configures the new firmware on the FPGA. The entire firmware upgrade process takes around 13 seconds to configure 10 RPC-DAQs. This paper explains the details of the architecture governing the firmware upgrade process and providing a comprehensive understanding of its mechanics.}


\begin{document}
\maketitle
\flushbottom

\section{Introduction}
\label{sec:intro}
\begin{figure}[htbp]
\centering
\includegraphics[width=.6\textwidth]{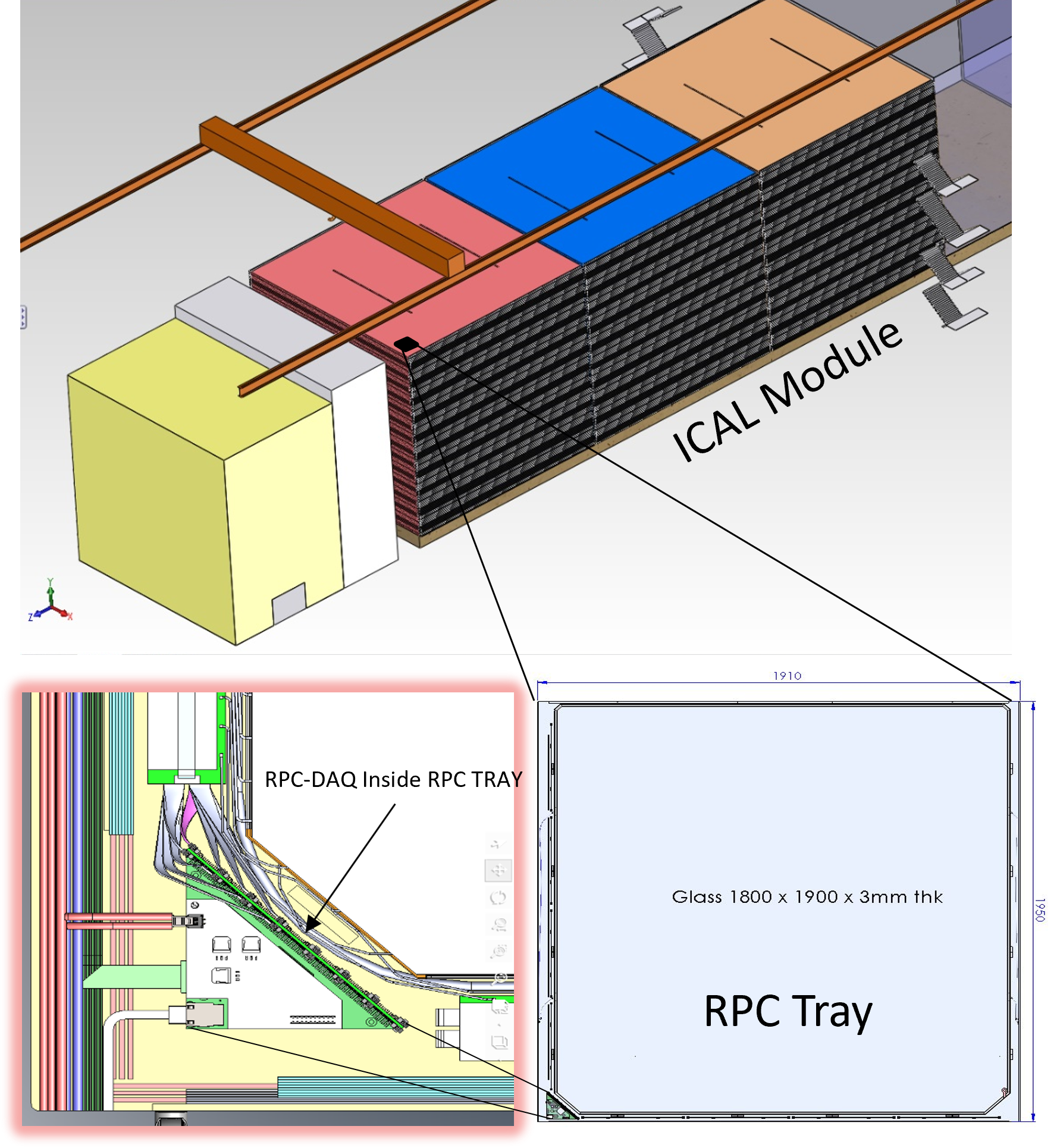}
\caption{RPC-DAQ position in INO-ICAL experiment. \label{fig:1}}
\end{figure}

In the context of the INO ICAL experiment~\cite{a}, where 28,800 RPCs are sandwiched in between stacked iron plates to observe muon events resulting from neutrino interactions, each RPC is provided with a front end electronics comprising of Analog Front End (AFE), Digital Front End (DFE) and High voltage supply (HV). The RPC signals produced are first amplified and converted to logic signals and sent to FPGA based Data Acquisition module (RPC-DAQs)~\cite{b} employed close to the RPC as DFE shown in Figure~\ref{fig:1} . The RPC-DAQ's FPGA firmware encompasses essential hardware logic to measure the position and timing information of incoming RPC signals. 

To support data transfer and command control RPC-DAQ is equipped with a Wiznet W5300 Ethernet Controller supporting TCP/IP. A softcore processor NIOS~\cite{c} is instantiated to offload DAQ tasks such as handling command execution, hardware readout, data transfer and detector health monitoring. Also to store the FPGA Configuration file an 8MB EPCS64 Flash Memory is interfaced with the FPGA. During installation and commissioning the RPC-DAQ's flash memory is configured with default factory firmware. Once installed the only way of accessing RPC-DAQ is through Ethernet~\cite{d}. A dedicated command interface using customized UDP protocol is implemented for modifying various DAQ and detector configuration. FPGA firmware upgrading is typically accomplished using JTAG programmers utilizing tools such as VIVADO Hardware Manager or Intel Quartus Prime Programmer for accessible DAQs. The traditional method of configuring FPGA firmware requires physical proximity to the detector. This poses a challenge given the large number of DAQs and their inaccessibility during the INO ICAL Experiment see Figure~\ref{fig:1}.

\begin{figure}[htbp]
\centering
\includegraphics[width=.5\textwidth]{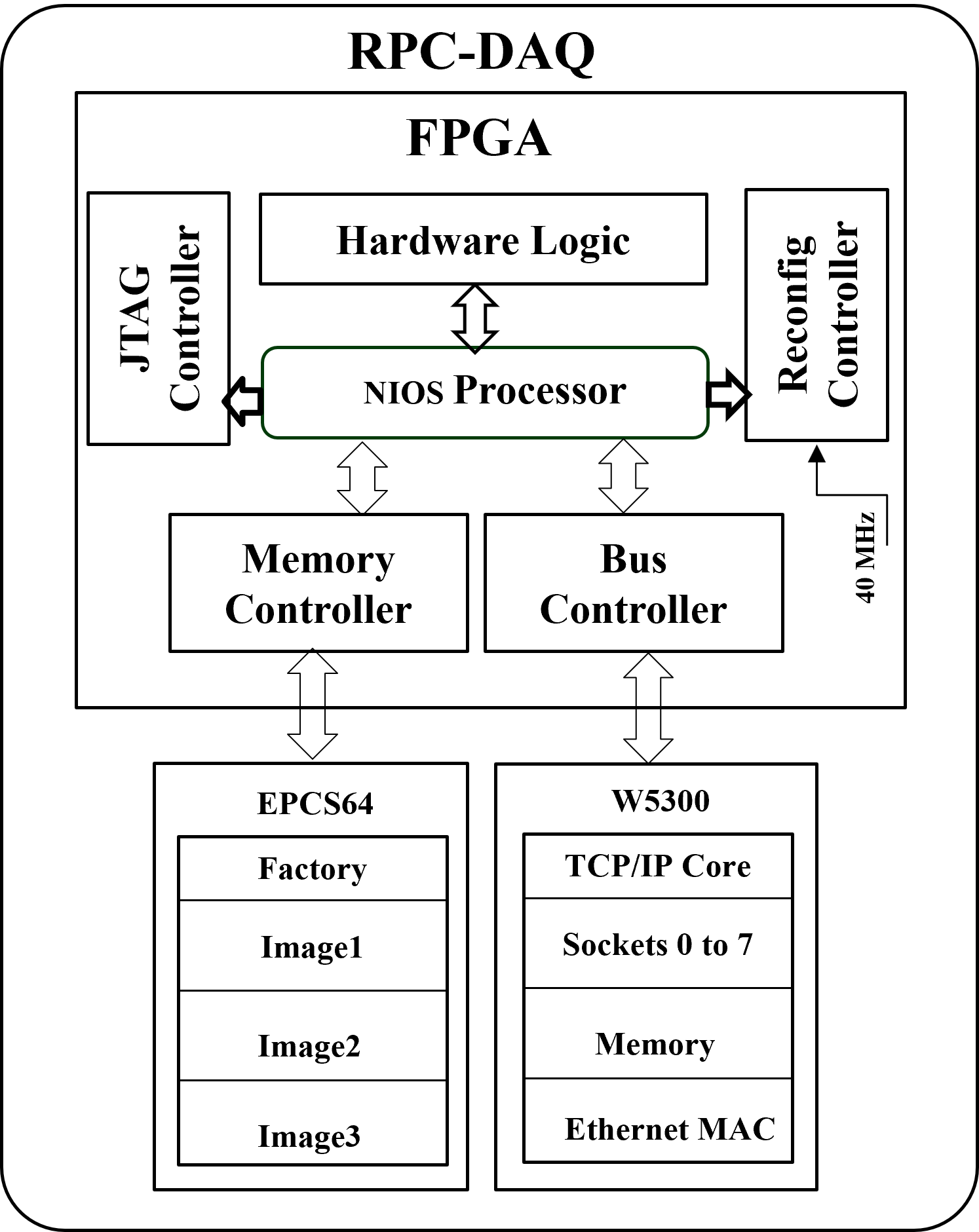}
\caption{RPC-DAQ Interface for Remote Firmware Upgradation.\label{fig:2}}
\end{figure}

Remote Firmware Upgradation (RFU) technique is commonly used to upgrade operating systems in mobile devices. This technique is well suitable in the INO ICAL experiment to configure the firmware of remotely located RPC-DAQs Figure~\ref{fig:1}. The instantiated NIOS processor in RPC-DAQ's Cyclone 4 FPGA~\cite{e} interfaces with the Wiznet W5300 Ethernet controller which uses sockets for command and real-time data transfer along with one dedicated socket for remote firmware upgrading. A dedicated back-end software "Remote System Upgrader (RS-Upgrader)" converts the FPGA raw binary firmware file into TCP packets and sends them to RPC-DAQs using their respective IP addresses. The NIOS Processor in the RPC-DAQ FPGA receives and overwrites the existing firmware in the flash memory as illustrated in Figure~\ref{fig:2}.

The standard configuration involves partitioning the flash memory to accommodate various firmware images alongside the factory firmware. The NIOS Processor interfaces with a reconfiguration Controller IP~\cite{f} as shown in Figure~\ref{fig:2} to trigger reconfiguration process. Users can load firmware image files into designated locations in the flash memory and switch between these firmware images or factory using the reconfiguration controller.  The factory firmware incorporates selection logic to facilitate the seamless choice of the required image. This paper describes various processes involved in the Remote firmware upgrade implementation in INO-ICAL experiment and its results. 

\section{Pre Requisites for Remote Firmware Upgradation}
\label{sec:pre_req}
To facilitate remote firmware upgradation in the INO ICAL experiment essential prerequisites must be met. These include the presence of FPGA-based Digital Systems (RPC-DAQs) with instantiated Soft or Hard Core Processor like (NIOS) and Ethernet controller like (Wiznet W5300)~\cite{g} along with correctly configured flash memory storing the initial factory Firmware. A reliable and stable Ethernet connection between RPC-DAQs and the back-end server is required. The RPC-DAQ firmware should support TCP/IP protocols.  Also a dedicated back-end software is required to convert the FPGA raw binary firmware files into TCP packets and establish a handshaking communication with the RPC-DAQ. A stable power supply and a controlled testing environment are required for uninterrupted operations.

\section{UDP and TCP Packet Formats used in RFU}
\label{sec:pack}
\begin{figure}[htbp]
\centering
\includegraphics[width=.8\textwidth]{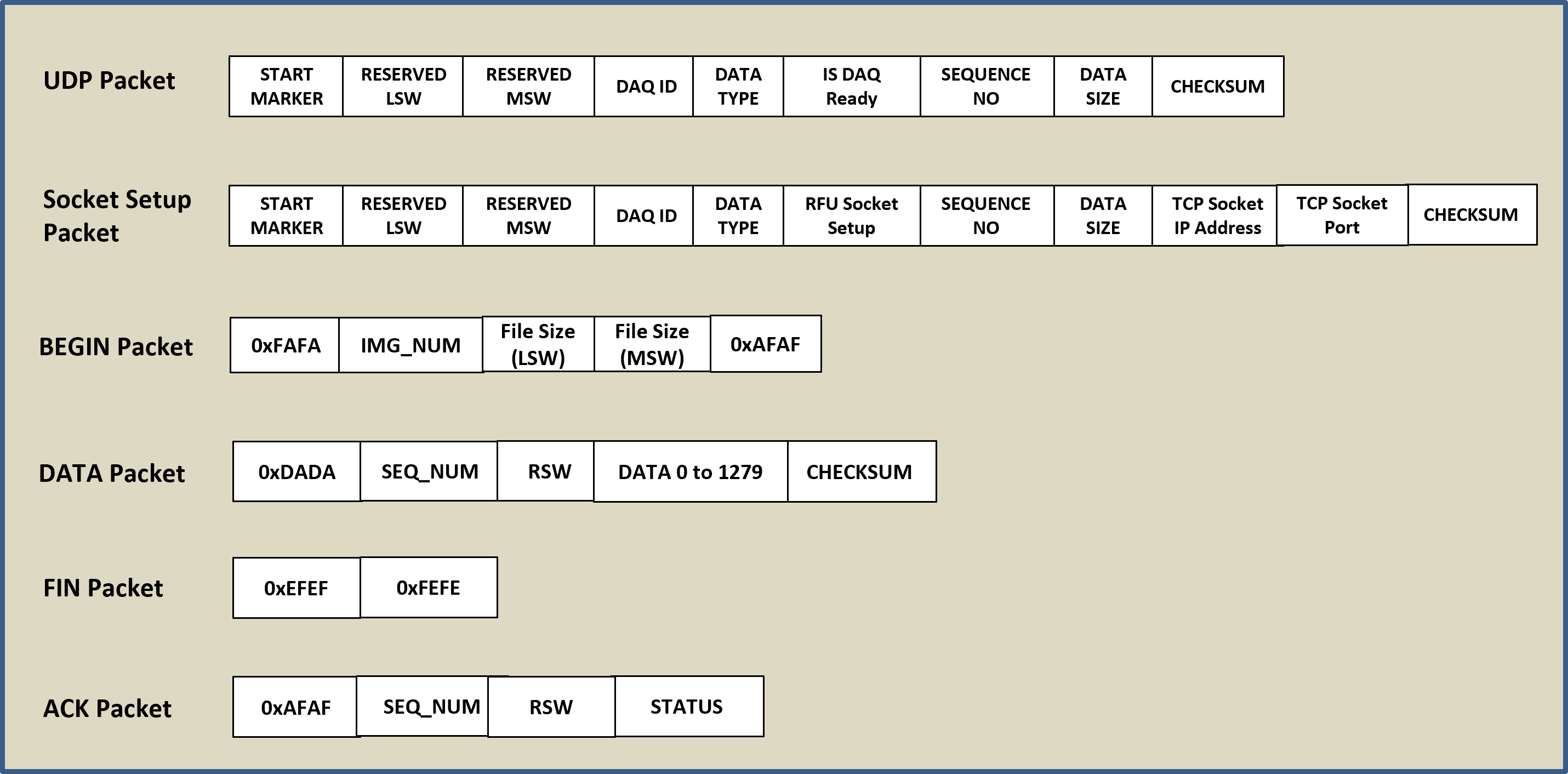}
\caption{Packet Formats between RPC-DAQ and back-end.\label{fig:3}}
\end{figure}

The remote firmware upgradation process follows several steps each initiated or controlled by UDP and TCP packets. The initial two packets illustrated in the Figure~\ref{fig:3} are UDP packets utilized by the command interface for retrieving RPC-DAQ status and configuring the TCP sockets in RPC-DAQ as server. The remote firmware upgrade process uses TCP protocols for both firmware transfer and acknowledgments. This file transfer process consist of three types of packets named as "BEGIN", "DATA" and "FIN" as depicted in the Figure~\ref{fig:3}. The BEGIN packet includes details about the firmware size and the address of the flash memory location where the firmware needs to be stored. DATA packets contain 1280 byte chunks of the firmware file as payload. The size 1280 byte was chosen optimally for TCP packet transmission. The FIN packet serves as an end marker signaling RPC-DAQ that all firmware data has been transferred. For each TCP packet a handshaking acknowledgement packet with the status (success or failure) of the previous process is employed. Each packet type used in remote firmware upgradation is characterized by distinct headers for identification purposes.

\section{Firmware Upgradation over LAN}
\label{sec:rfu}
The connection between the RPC-DAQs and the back-end RS-Upgrader is established through an existing robust and efficient connectionless UDP command interface namely Hybrid Protocol Command Interface (HPCI)~\cite{d}. In Remote Firmware Upgradation, HPCI enables the RPC-DAQ as server and the RS-Upgrader as the client. In scenarios involving simultaneous upgrades of multiple RPC-DAQs the network operates in a 'one client and many servers' topology. Upon receiving the "RFUSOCKSETUP" UDP command from the back-end, participating RPC-DAQs prepare for system upgrade and establish a TCP connection. The back-end initiates the firmware transfer as TCP packets whereas RPC-DAQ receives each of these firmware packets and writes directly into the flash memory at a user-defined location. For every TCP firmware packet received, an acknowledgement is promptly sent.

\begin{figure}[htbp]
\centering
\includegraphics[width=.8\textwidth]{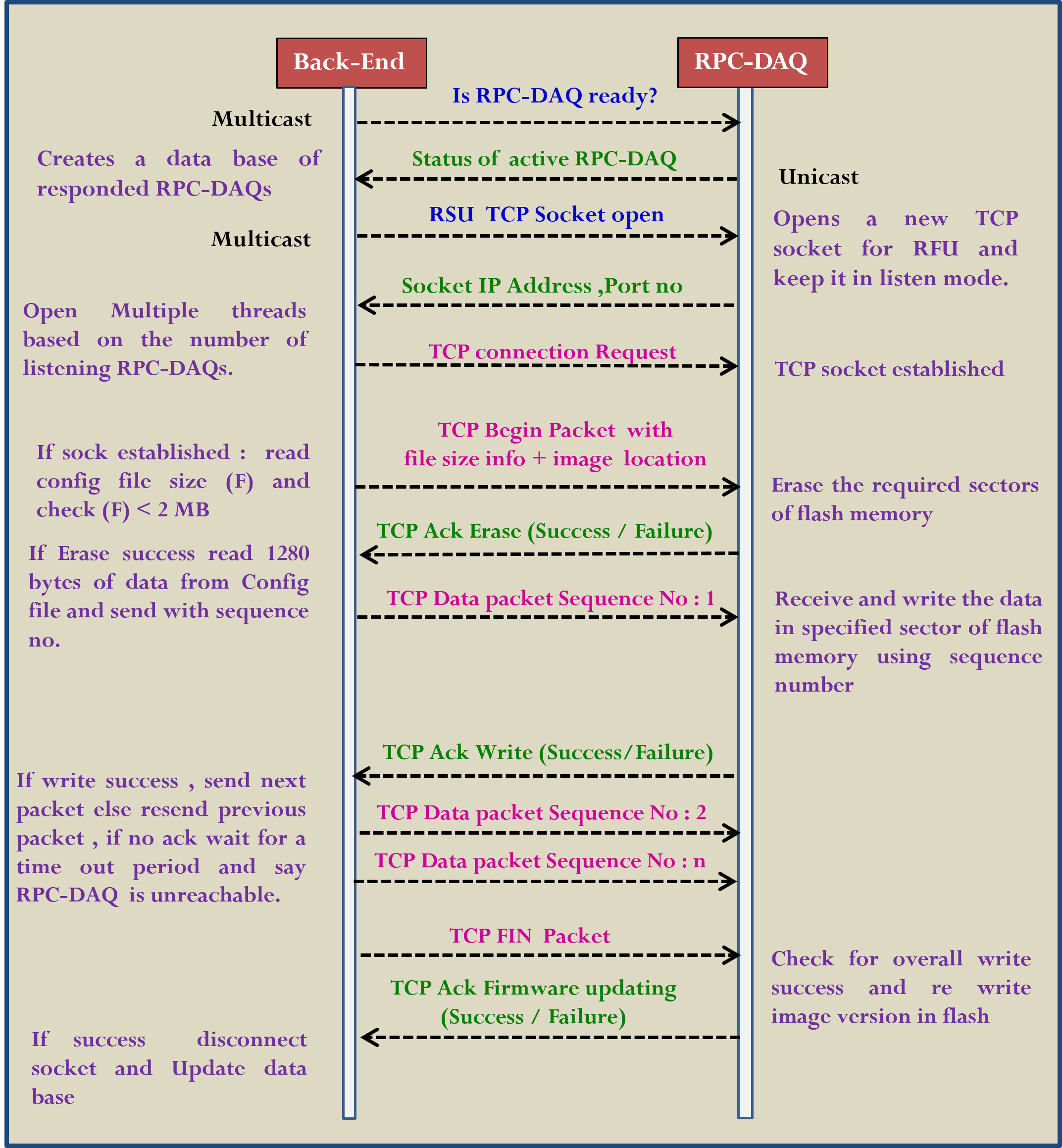}
\caption{Remote Firmware Upgradation steps involved.\label{fig:4}}
\end{figure}

This handshake based protocol eliminates the risk of misconfigurations in the RPC-DAQs. The back-end client manages each firmware and acknowledgement packet using proper sequence numbers reducing packet losses through retransmission. RPC-DAQs can access any location in their flash memory for loading firmware allowing use of multiple firmware versions. During the configuration of multiple RPC-DAQs the RS-Upgrader employs multiple threads each handling a few sockets with dedicated port numbers for each socket. All network protocols include a user checksum on both ends to ensure data integrity. The time required for writing and acknowledging one packet depends on the flash memory's write time. Timeout values are set based on the total cycle time of a TCP firmware packet. In case of a checksum failure or timeout, the retransmission process is triggered. Figure~\ref{fig:4} illustrates the comprehensive communication scheme between RPC-DAQs and the RS-Upgrader ensuring a smooth and secure firmware upgrading process.

\subsection{BEGIN Packet}
\label{sec:begin}
Upon receiving the 'BEGIN' packet from the back-end the RPC-DAQ NIOS processor starts erasing the firmware image in the flash memory at the specified location. Erasing the flash memory sector by sector takes upto a few seconds. The back-end server is kept in waiting state until a positive acknowledgment from RPC-DAQs arrive.

\subsection{DATA Packet}
\label{sec:data}
After successful 'BEGIN' packet acknowledgment each thread dedicated to each RPC-DAQ starts framing Data Packets of size 1280 bytes. Using the dedicated TCP port number the back-end send firmware 'DATA' packets to its respective RPC-DAQ. To enhance reliability, each DATA packet is equipped with an XOR checksum word both to and fro. Each data packet is provided with sequence number and the same number is retransfered in the acknowledgment packet. The back-end uses this sequence number correctly to maintain 'DATA' packet sequence.

\subsection{FIN Packet}
\label{sec:fin}
When the last Firmware Packet is acknowledged the back-end software sends a 'FIN' packet to indicate the end of the upgrade process. On receiving the FIN (termination) packet, RPC-DAQs close the active socket and return to its main loop.

\section{Flash Memory Access by NIOS}
\label{sec:flash}

\begin{figure}[htbp]
\centering
\includegraphics[width=.6\textwidth]{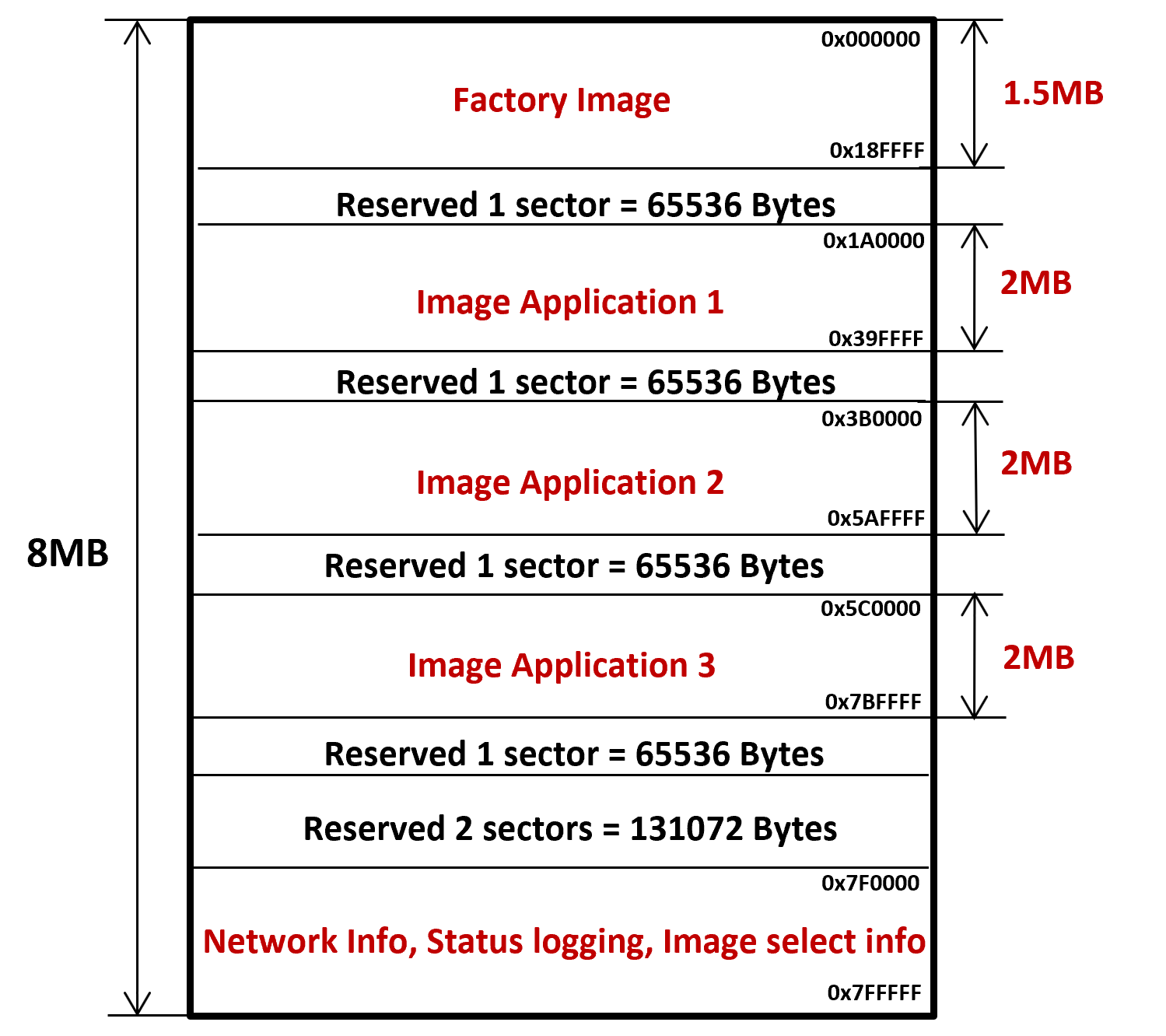}
\caption{FLash Memory Breakin Diagram.\label{fig:5}}
\end{figure}

The Remote firmware Upgradation functionality allows the loading of Flash Memory within RPC-DAQs with up to three user firmware versions and one factory firmware each with a size less than 2MB. This strategic allocation of firmware images is illustrated in Figure~\ref{fig:5}. The EPCS64 Flash memory~\cite{h} has a capacity of 64 Megabits or 8 Megabytes, organized into sectors with each sector having a depth of 65536 Bytes. To accommodate multiple firmware versions a firmware size of 2MB was selected and one sector is reserved between each image firmware sector as boundary. The final sector is exclusively designated for storing DAQ-specific information such as network configuration details (MAC, IP, GATEWAY), and boot image information. The factory firmware image functions as a boot loader. During the boot sequence of the factory firmware the NIOS processor reads the boot image information from the last sector of the flash memory and loads the reconfiguration controller to initiate the booting process for the image firmware. It is noteworthy that during installation and commissioning the RPC-DAQ flash memory will initially contain only the factory firmware devoid of any image firmware's. Therefore users must first write image firmware's into the flash memory before triggering a reboot.

\section{Back-end Networking Application Software}
\label{sec:Back_soft}
\begin{figure}[htbp]
\centering
\includegraphics[width=.6\textwidth]{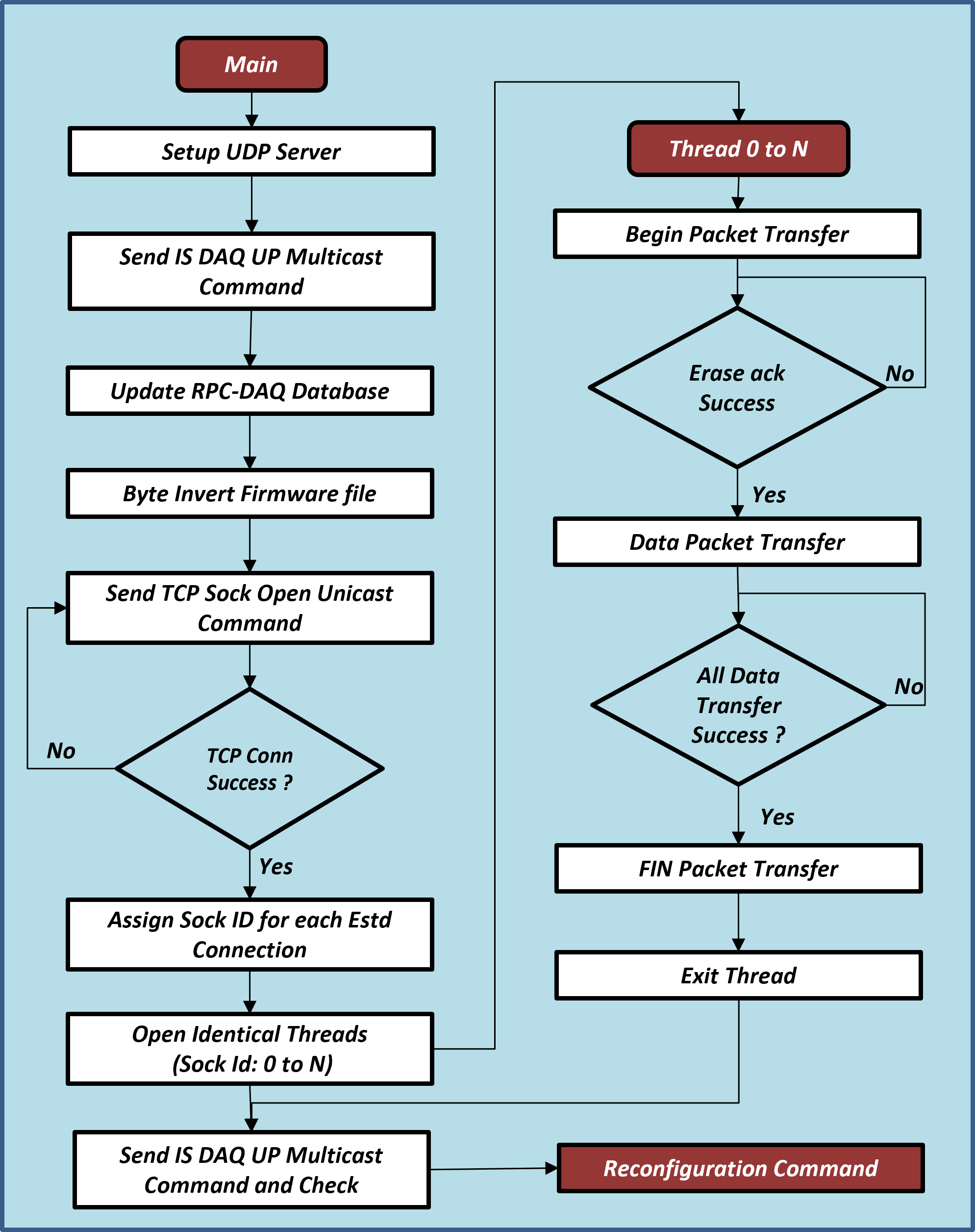}
\caption{Back-end Software Flowchart.\label{fig:6}}
\end{figure}

A C/C++ based back-end software namely RS-Upgrader has been designed to control the remote upgradation process. This software uses fundamental socket programming principles to establish both UDP and TCP sockets. The software also maintains a local database containing information about all available RPC-DAQs. The algorithm incorporates a multi-threading feature capable of handling up to 20 RPC-DAQs concurrently. Illustrated in the Figure~\ref{fig:6} the main loop is responsible for preparing the firmware configuration file and establishing connections with RPC-DAQs. The thread handles the transmission of firmware files to RPC-DAQs and handles acknowledgments.

\section{FPGA Reconfiguration Procedure}
\label{sec:Reconfig}

\begin{figure}[htbp]
\centering
\begin{subfigure}[c]{0.5\textwidth}
\includegraphics[width=\linewidth]{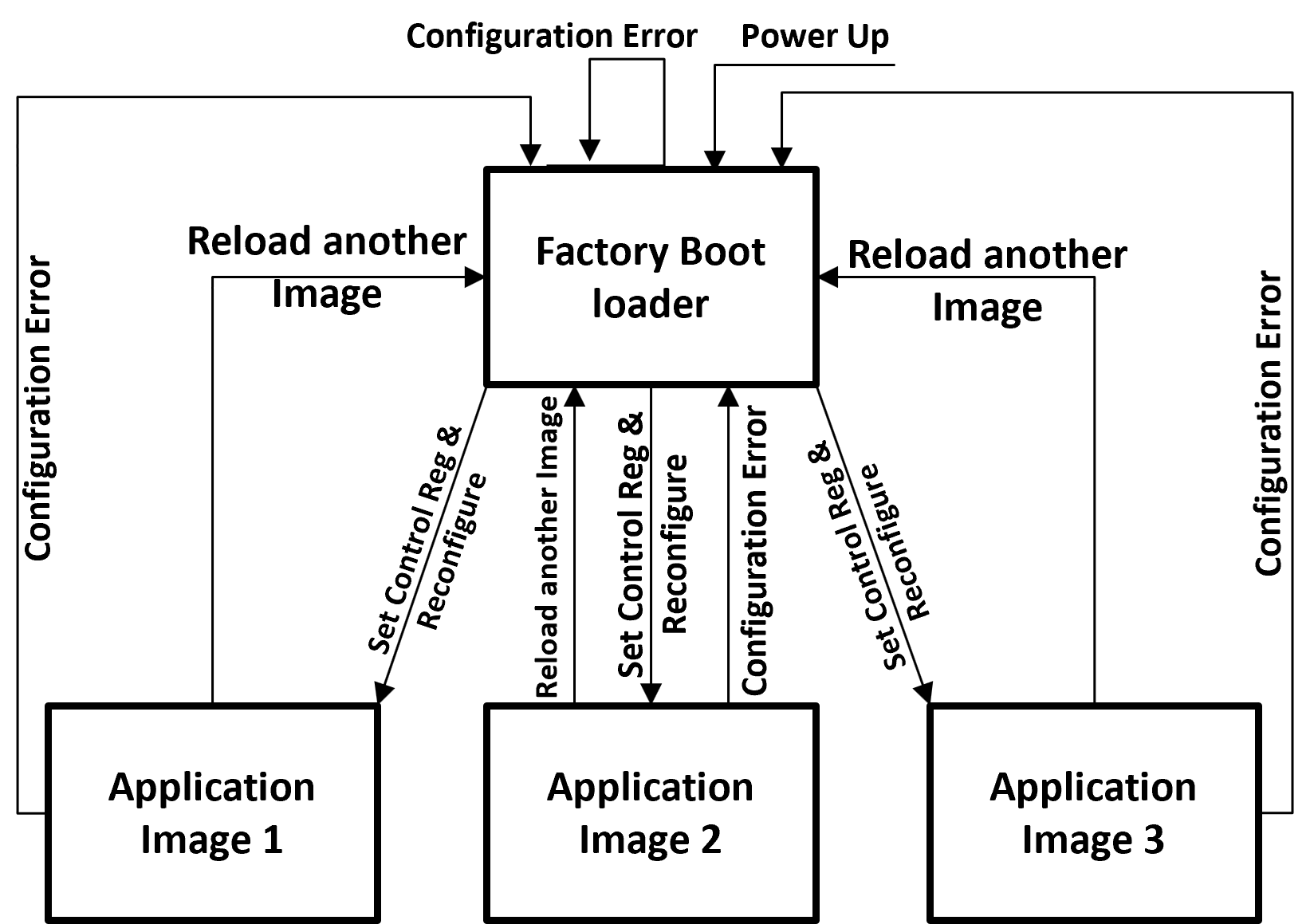} 
\caption{Reconfiguration Controller logic and fail safe.}
\label{fig:7a}
\end{subfigure}\hfill  
\begin{subfigure}[c]{0.4\textwidth}
\includegraphics[width=\linewidth]{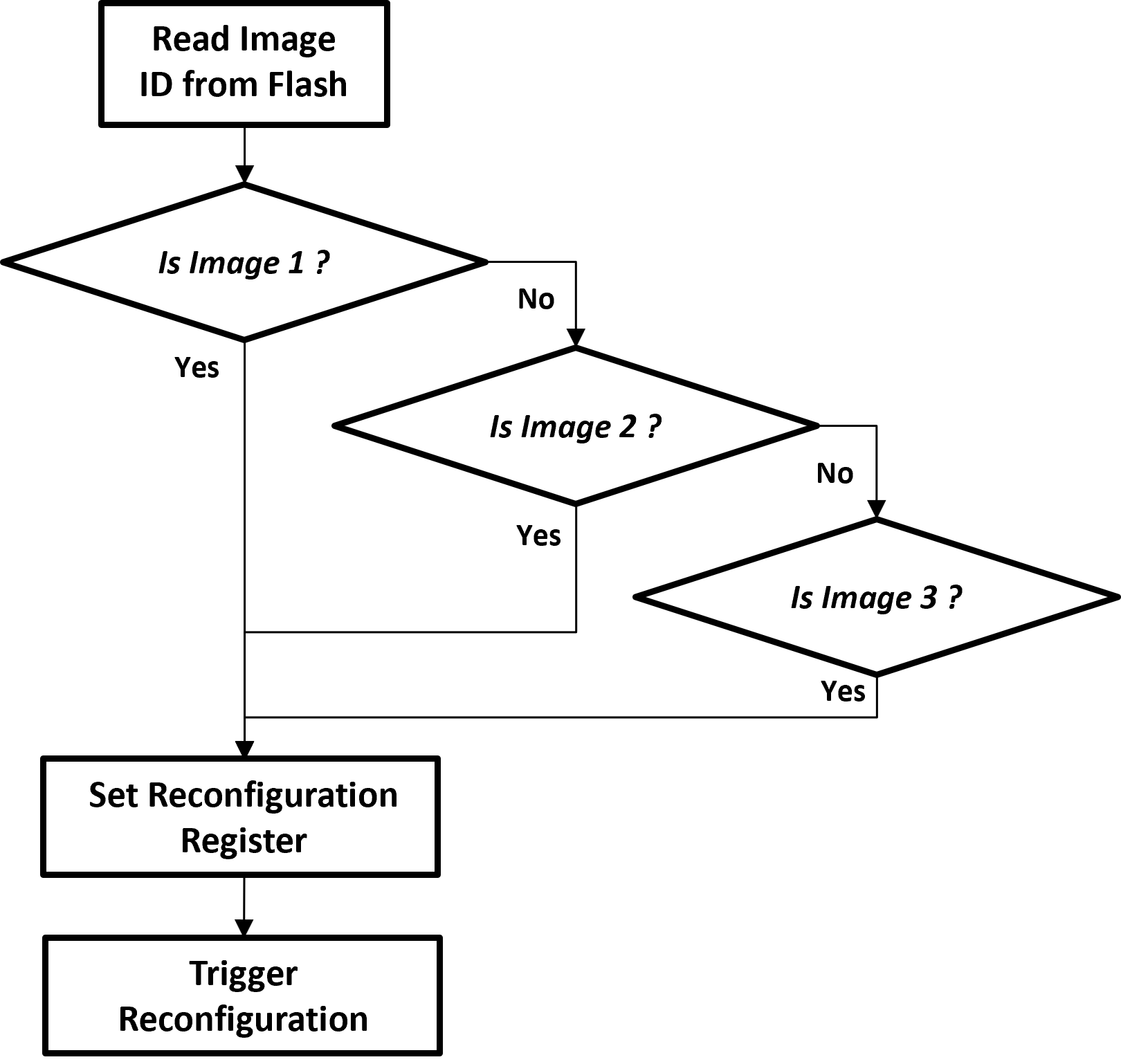}
\caption{Image selection during factory booting.}
\label{fig:7b}
\end{subfigure}
\caption{Factory and Image Reconfiguration by the Boot Loader. \label{fig:7}}
\end{figure}

Upon power-up the FPGA initiates the boot sequence with the factory firmware. In this sequence the factory firmware reads the image number from the last sector of the flash memory. Based on its value (01/02/03) the factory firmware sets the base sector address of the image in the reconfig register of the reconfiguration controller and initiates the reconfiguration process as illustrated in the Figure~\ref{fig:7}(a). A dedicated UDP command is utilized to modify the image number in the flash memory. In case the written image firmware is corrupted due to packet integrity issues during the transfer the reconfiguration controller halts the reconfiguring process, issues a configuration error and reboots the factory firmware as shown in Figure~\ref{fig:7}(b). In addition to reconfiguring from the factory firmware a distinct UDP command has been implemented to directly trigger the reconfiguration process.


\section{Performance Studies and Results}
\label{sec:result}
\begin{figure}[htbp]
\centering
\begin{subfigure}[c]{0.4\textwidth}
\includegraphics[width=\linewidth]{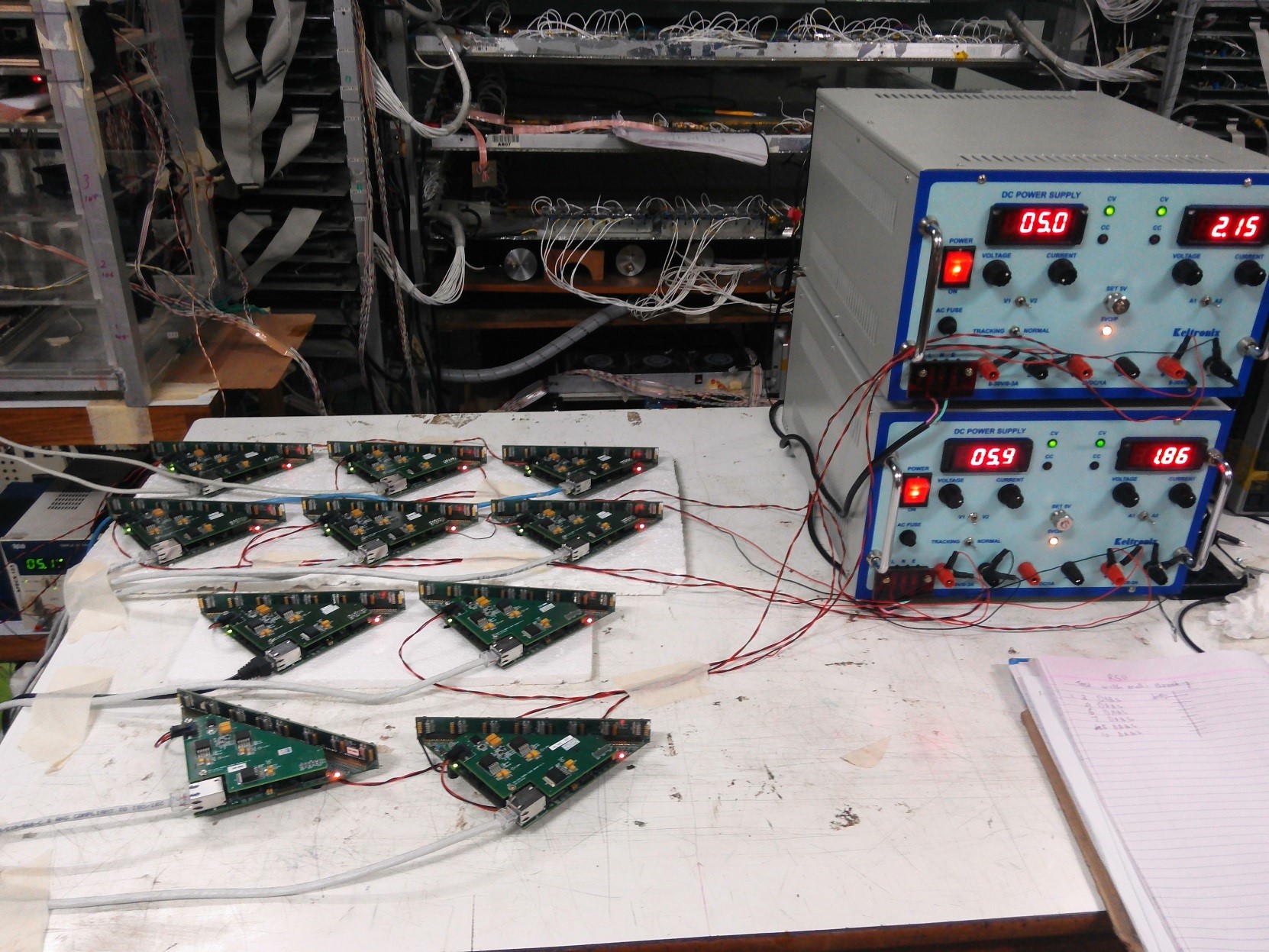} 
\caption{RPC-DAQ Remote firmware upgrade test setup.}
\label{fig:8a}
\end{subfigure}\hfill  
\begin{subfigure}[c]{0.5\textwidth}
\includegraphics[width=\linewidth]{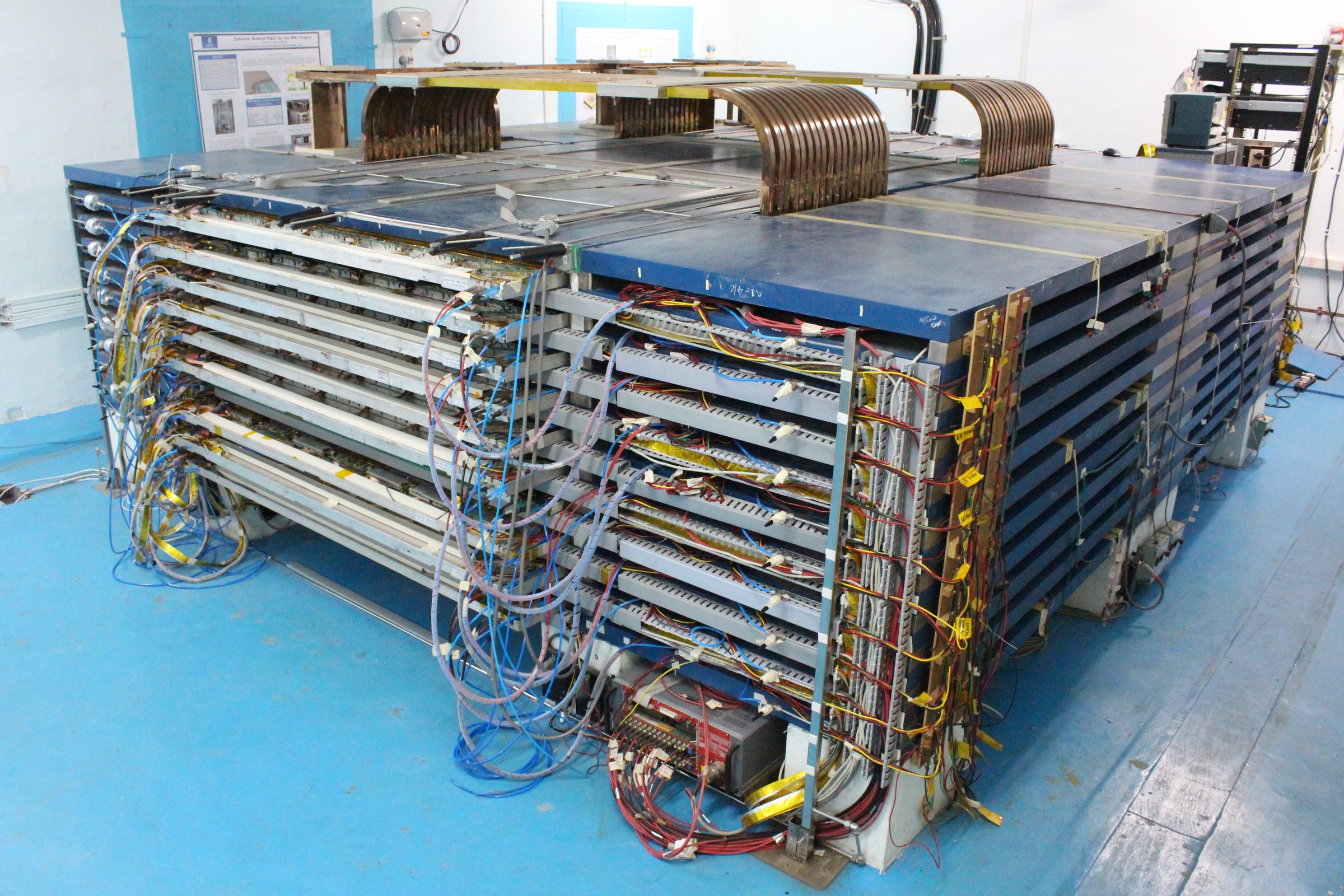}
\caption{Mini-ICAL Detector with RPC and RPC-DAQs installed.}
\label{fig:8b}
\end{subfigure}
\caption{Testing of Remote Firmware Upgradation on test bench and installation in Mini-ICAL Detector.\label{fig:8}}
\end{figure}

To assess the functionality of remote firmware upgrading a test bench is set up with 10 RPC-DAQs each equipped with power supply and Ethernet interface as illustrated in Figure~\ref{fig:8}(a). Factory and three dummy image firmwares are generated each with distinct logic to control two LEDs exhibiting different LED states outlined in Table~\ref{tab:1}. 


This setup allows verification of the proper configuration of the new image after firmware updates. Also to test the boot failures, we transferred only half firmware data into flash memory image and try to boot from this image. The reconfiguration controller stops the reconfiguration and returns to factory firmware. The efficiency of remote firmware upgrades is measured by the time required to upgrade an RPC-DAQ firmware. The upgrade process involves setting up the RPC-DAQ with a TCP connection erasing the old firmware image from the flash memory and writing the new firmware image into the flash memory. The duration of these three processes is detailed in Table~\ref{tab:2}.

\begin{table}[htbp]
\centering
\caption{LED states Vs Image Firmware.\label{tab:1}}
\smallskip
\begin{tabular}{c|c|c}
\hline
Firmware  &LED1 &LED2\\
\hline
Factory & Low & Low\\
Image1 & Low & High\\
Image2 & High & Low\\
Image3 & High & High\\
\hline
\end{tabular}
\end{table}

The Reconfiguration Controller takes approximately $1\, \text{ms}$ to load a new image into the FPGA. It's important to adjust the watchdog timer logic in the hardware during the time consuming image writing process in the flash memory. After successfully verifying the remote firmware upgrading process in the test bench the RPC-DAQ modules are installed in the prototype mini ICAL detector~\cite{i,j} shown in the Figure~\ref{fig:8}(b), where the remote firmware update functionality is realistically tested with new firmwares. As depicted in the Figure~\ref{fig:8} accessing RPC-DAQs post installation in mICAL is practically impossible. The implementation of Remote Firmware Upgradation has been consistently successful in mICAL since 2018 without encountering any failures.

\begin{table}[htbp]
\centering
\caption{Time Duration of each Process in Remote Firmware Upgradation.\label{tab:2}}
\smallskip
\begin{tabular}{l|c}
\hline
Process & Time Taken (Seconds)\\
\hline
Initialization & 0.03 \\
Erasing Flash memory & 8  \\
Writing new Firmware & 13  \\
\hline
\end{tabular}
\end{table}
\section{Conclusion}

The Remote Firmware Upgrade system has been successfully implemented, commissioned and used in the current mICAL detector. Presently the process allows configuration for only 10 RPC-DAQs simultaneously. In the ICAL Experiment the objective is to concurrently configure one hundred RPC-DAQs aiming for increased flexibility. This goal can be realized by employing additional CPU cores and maximizing network bandwidth. The remote firmware upgrade methodology presented in this paper can be adapted to any data acquisition system equipped with an FPGA processor or microcontroller, flash memory, and an Ethernet controller. To achieve higher throughput for remote firmware upgrading an increase in the processor clock from 50 MHz to higher values may be considered.


\acknowledgments
We sincerely thank all our present and former INO colleagues especially Puneet Kanwar Kaur, Umesh L, Upendra Gokhale, Anand Lokapure, Suraj Kole, Sagar Sonavane, Salam Thoi Thoi, Rajkumar Bharathi for technical support during the design, Also we would like to thank members of TIFR, namely S.R. Joshi, Piyush Verma, Darshana Gonji, Santosh Chavan and Vishal Asgolkar who supported testing and commissioning. Also we like to thank former INO directors N.K. Mondal and V. M Datar, for their continuous encouragement and guidance.


\end{document}